\documentclass[11pt]{article}
\usepackage{fullpage}
\usepackage{adjustbox}
\usepackage{xspace}
\usepackage{lmodern}
\usepackage{amsfonts}
\usepackage{hyperref}
\usepackage{amssymb}
\usepackage{xcolor}
\usepackage{amsmath}
\usepackage{cleveref}
\usepackage{subcaption}
\usepackage{textgreek}

\usepackage[utf8]{inputenc}

\usepackage{listings}

\definecolor{primary}{RGB}{0, 51, 102}    
\definecolor{secondary}{RGB}{230, 240, 255} 
\definecolor{accent}{RGB}{100, 100, 100}  

\definecolor{keywordcolor}{rgb}{0.7, 0.1, 0.1}   
\definecolor{tacticcolor}{rgb}{0.0, 0.1, 0.6}    
\definecolor{commentcolor}{rgb}{0.4, 0.4, 0.4}   
\definecolor{symbolcolor}{rgb}{0.0, 0.1, 0.6}    
\definecolor{sortcolor}{rgb}{0.1, 0.5, 0.1}      
\definecolor{attributecolor}{rgb}{0.7, 0.1, 0.1} 

\usepackage{tikz}
\usepackage[edges]{forest}
\usetikzlibrary{shadows}

\usepackage{todonotes}

\usepackage{stmaryrd}   
\usepackage{graphicx}   
\usepackage{datetime}   
\usepackage{times}
\usepackage{comment}
\usepackage{floatflt}
\usepackage{wrapfig}

\usepackage[framemethod=TikZ]{mdframed}
\usepackage{xspace}
\usepackage{authblk}

\newcommand{\cslib}{CSLib\xspace}

\title{\cslib: The Lean Computer Science Library}
\author[1,2]{Clark Barrett}
\author[3,4]{Swarat Chaudhuri}
\author[5]{Fabrizio Montesi}
\author[1]{Jim Grundy}
\author[3]{Pushmeet Kohli}
\author[6,1]{Leonardo de Moura}
\author[7,8]{Alexandre Rademaker}
\author[9]{Sorrachai Yingchareonthawornchai}

\affil[1]{Amazon}
\affil[2]{Stanford University}
\affil[3]{Google DeepMind}
\affil[4]{The University of Texas at Austin}
\affil[5]{FORM, University of Southern Denmark}
\affil[6]{Lean FRO}
\affil[7]{Atlas Computing}
\affil[8]{Getulio Vargas Foundation}
\affil[9]{Institute for Theoretical Studies, ETH Zürich}
\date{}

\begin{document}
\maketitle

\begin{abstract}
We introduce \cslib, an open-source framework for proving computer-science-related theorems and writing formally verified code in the Lean proof assistant.
\cslib aims to be for computer science what Lean’s Mathlib is for mathematics. Mathlib has been tremendously impactful: it is a key reason for Lean's popularity within the mathematics research community, and it has also played a critical role in the training of AI systems for mathematical reasoning. However, the base of computer science knowledge in Lean is currently quite limited. \cslib will vastly enhance this knowledge base and provide infrastructure for using this knowledge in real-world verification projects. By doing so, \cslib will (1) enable the broad use of Lean in computer science education and research, and (2) facilitate the manual and AI-aided engineering of large-scale formally verified systems.
\end{abstract}

\section{Introduction}

On its face, Mathlib \cite{mathlibtechrep} is a library of mathematical results and techniques formalized in the Lean \cite{moura2021lean} proof assistant. In reality, it is much more than that---it is a community effort by mathematicians all over the world to collect the world's mathematical knowledge in a form that can easily be vetted and shared.  Mathlib provides a platform for a new way to do mathematics, where many people can confidently collaborate on large and challenging projects \cite{ringer2024proofs}. Also, by representing a wide range of mathematical definitions and proofs in a machine-checkable form, it opens up new opportunities for AI systems to aid the process of mathematical discovery \cite{yang2025position,alphaproof2025}. This new approach to mathematics is being embraced by a new generation of mathematicians and has gained many prominent champions \cite{tao2024machine,kontorovich2025shape}. Mathlib is at the heart of this paradigm shift, and as such, it is hard to overstate its impact. 

Computer science has always been close to mathematics.  Many computer science departments originally emerged as offshoots of mathematics departments, and even today, the distinction between computer science theory and mathematics is blurry.  Moreover, computer scientists have long recognized formal mathematics as a power tool that enables the design of reliable and secure systems \cite{boyer1983proof,clarke1997model,hoare2003verifying}. However, until now there has been no effort to build a Mathlib-like universal repository of formalized computer science knowledge.

In this white paper, we introduce \cslib,\footnote{The main \cslib website is at \url{https://cslib.io}. The current \cslib codebase is at \url{https://github.com/leanprover/cslib/}.} a library for Lean-based formal verification and computer science research that seeks to change this state of affairs. \cslib has two ``pillars":
\begin{enumerate}
\item \emph{Formalizing all essential computer science concepts in Lean}. The formalizations include, but are not limited to, models of computation such as the $\lambda$-calculus and resource-bounded Turing machines; numerous algorithms and data structures accompanied by proofs of correctness and complexity; concurrency theory; foundations of programming languages; and mathematical tools relevant to computer science that do not currently appear in Mathlib.
\item \emph{Building an infrastructure for Lean-based reasoning about everyday imperative code}. This infrastructure consists of an intermediate programming language, called Boole, that allows classical imperative constructs to be interspersed with specifications written in Lean; machinery for generating Lean-language verification conditions from Boole code; and verified Boole implementations of the algorithms and data structures covered in Pillar 1.
\end{enumerate}

Collectively, the Lean code and proofs in \cslib enable the use of Lean as a medium for mathematically grounded computer science research and development. For example, we can imagine a theory researcher using \cslib's built-in abstractions to quickly mock up a novel approximation algorithm and theorems capturing its essential properties, a programming languages researcher using the framework to build verified compilers, or a systems researcher using the library to model new network protocols with worst-case guarantees. The Boole framework offers a crucial bridge from Lean to everyday programming languages. Specifically, we envision a world in which code in 
mainstream languages like Rust and C++ is automatically translated into Boole at scale. 
\cslib's built-in infrastructure turns the task of verifying these Boole-language programs into that of proving Lean-language theorems. This proof task is then solved using Lean-language proof machinery.

\paragraph{\cslib's Impact.} 
We see the impact of \cslib taking two forms: addressing an urgent practical need for reliable software systems, and fulfilling a long-term need for more rigorous and scalable computer science.

On the first point, progress on software reliability and security is more time-critical than ever because of new risks that developments in AI pose to the world's computing infrastructure.  Today's AI systems have a remarkable ability to analyze code, making it much easier for an AI-powered malicious actor to find and exploit vulnerabilities.  At the same time, AI coding agents are writing more and more of the world's code. While this leads to overall productivity gains, AI can also introduce correctness errors and security vulnerabilities into code \cite{perry2023users}. Formal methods can mitigate these risks by 
{mathematically proving} code---whether human-written or AI-generated---to be free of bugs and vulnerabilities.

However, the impact of formal methods has been historically constrained by the immense human effort that they require. 
Consider seL4 \cite{klein2009sel4}, the world's first formally verified operating system kernel. 
The system had a profound impact: an seL4-powered drone in the DARPA HACMS program famously resisted six weeks of sustained attacks by an elite red team, leading the drone to be called the ``most secure drone in the world.'' But seL4 also highlights the cost penalties associated with traditional formal verification. The system required over 20 person-years of proof effort and included about 480,000 lines of formal specification and proof in the Isabelle \cite{nipkow2002isabelle} framework. Such costs are difficult to justify in most commercial settings. 


The \cslib vision addresses this bottleneck in two ways. Once \cslib has been built, a developer would be able to derive reliable and secure systems compositionally, by putting together pre-verified \cslib components. Second, \cslib can simplify the use of AI in formal verification. AI systems have already demonstrated superhuman formal proof capabilities in the pure mathematics setting  \cite{alphaproof2025}. However, to leverage these capabilities in broad-domain code verification, we need a comprehensive language in which to specify code properties. Lean and \cslib will provide this language. In addition, \cslib will serve as a repository of high-quality data on which to train AI systems for formalization and formal proof.

Beyond this impact on practical formal verification, \cslib will serve as a platform for 21st-century computer science research. The formal verification that \cslib enables will induce confidence in claims and allow the large-scale reuse of foundational concepts. We can also imagine AI tools trained on \cslib discovering novel algorithms and settling open claims, extending to computer science the role that AI agents are beginning to play in mathematics research. 

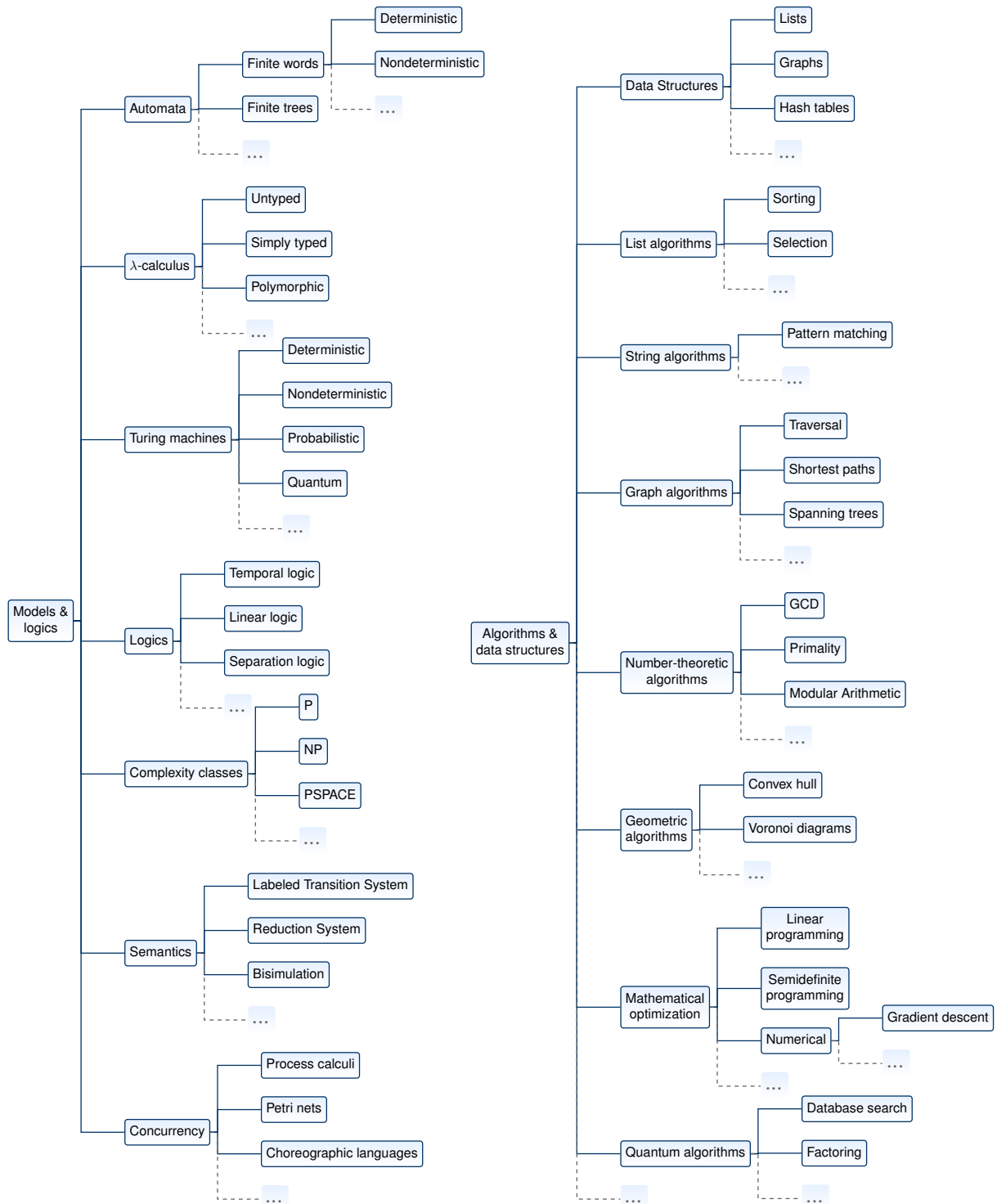
\begin{figure}
\centering
\begin{subfigure}{0.45\linewidth}
\begin{adjustbox}{max height=0.9\textheight, keepaspectratio}
\begin{forest}
  for tree={
    align=center, 
    grow'=east,               
    draw=primary,
    thick,
    rounded corners=2pt,
    top color=secondary,
    bottom color=white,
    font=\sffamily\small,
    edge={draw=primary, thick},
    l sep+=25pt,             
    s sep+=6pt,              
    anchor=west,
    child anchor=west,
    parent anchor=east,
    forked edges,   
  },
  ellipsis/.style={
    draw=none, fill=none, 
    font=\bfseries\color{accent},
    edge={draw=accent, dashed}, l sep=10pt
  }
  [Models \& \\ logics 
    [Automata
      [Finite words
        [Deterministic]
        [Nondeterministic]
        [\dots, ellipsis]
      ]
      [Finite trees]
      [\dots, ellipsis]
    ]
    [$\lambda$-calculus
      [Untyped]
      [Simply typed]
      [Polymorphic]
      [\dots, ellipsis]
    ]
    [Turing machines
      [Deterministic]
      [Nondeterministic]
      [Probabilistic]
      [Quantum]
      [\dots, ellipsis]
    ]
    [Logics 
      [Temporal logic]
      [Linear logic]
      [Separation logic]
      [\dots, ellipsis]
    ]
    [Complexity classes
      [P]
      [NP]
      [PSPACE]
      [\dots, ellipsis]
    ]
    [Semantics
      [Labeled Transition System]
      [Reduction System]
      [Bisimulation]
      [\dots, ellipsis]
    ]
    [Concurrency
      [Process calculi]
      [Petri nets]
      [Choreographic languages]
      [\dots, ellipsis]
    ]
  ]
\end{forest}
\end{adjustbox}
\caption{Models \& logics}\label{fig:pillar1-a}
\end{subfigure}
\begin{subfigure}{0.45\linewidth}
\begin{adjustbox}{max height=0.9\textheight, keepaspectratio}
\begin{forest}
  for tree={
    align=center,
    grow'=east,
    draw=primary,
    thick,
    rounded corners=2pt,
    top color=secondary,
    bottom color=white,
    font=\sffamily\small,
    edge={draw=primary, thick},
    l sep+=25pt,
    s sep+=6pt,
    anchor=west,
    child anchor=west,
    parent anchor=east,
    forked edges,
  },
  ellipsis/.style={
    draw=none, fill=none, 
    font=\bfseries\color{accent},
    edge={draw=accent, dashed}, l sep=10pt
  }
  [Algorithms \& \\ data structures
    [Data Structures
      [Lists]
      [Graphs]
      [Hash tables]
      [\dots, ellipsis]
    ]
    [List algorithms
     [Sorting]
     [Selection]
      [\dots, ellipsis]
    ]
    [String algorithms
       [Pattern matching]
       [\dots, ellipsis]
    ]
    [Graph algorithms
      [Traversal]
      [Shortest paths]
      [Spanning trees]
      [\dots, ellipsis]
    ]
    [Number-theoretic \\ algorithms
      [GCD]
      [Primality]
      [Modular Arithmetic]
      [\dots, ellipsis]
    ]
    [Geometric \\algorithms
      [Convex hull]
      [Voronoi diagrams]
      [\dots, ellipsis]
    ]
    [Mathematical\\ optimization
    [Linear \\programming]
    [Semidefinite \\programming]
    [Numerical 
       [Gradient descent]
       [\dots, ellipsis]
      ]
    [\dots, ellipsis]
    ]
    [Quantum algorithms
      [Database search]
      [Factoring]
      [\dots, ellipsis]
    ]
    [\dots, ellipsis]
  ]  
\end{forest}
\end{adjustbox}
\caption{Algorithms and data structures}\label{fig:pillar1-b}
\end{subfigure}
\caption{\cslib's Pillar 1: A possible organization of covered topics.}\label{fig:pillar1}
\end{figure}

\section{Technical Approach}

Now we provide some more details on the two pillars of \cslib's technical approach. 

\subsection{Pillar 1: Formalizing Computer Science in Lean}\label{sec:Pillar1}

\cslib's first development pillar uses Lean to formalize a comprehensive body of models of computation, algorithms, and data structures, along with properties of these artifacts and proofs of these properties. The formalizations are designed to form a coherent and integrated framework rather than a collection of unrelated modules. Such unification leads to a whole that is greater than the sum of its parts and has been key to Mathlib's impact. 

The pillar has a synergistic relationship with Mathlib. On the one hand, it heavily uses Mathlib modules for, say, big-O reasoning and probability theory. On the other hand, it formalizes some core mathematics---for example, certain lemmas from combinatorics needed to analyze approximation algorithms or inequalities needed to prove the convergence of optimization algorithms---that are useful in computer science but do not currently appear in Mathlib. Depending on the circumstances, we may either contribute these formalizations to Mathlib or keep them within the \cslib repository. 

Now we provide some more details on this pillar's targets. 

\paragraph{Models \& Logics.} Over the years, computer scientists have designed a wide range of \emph{models of computation}. These include models for deterministic, nondeterministic, probabilistic, and quantum computation; models for functional computations, online algorithms, and interactive protocols; models that solve decision problems; and models that approximately solve optimization problems. Many of these models have never been formalized; the ones that have tend to exist in isolated repositories. A key goal of \cslib's Pillar 1 is to create a unified body of formalizations of these models. 

Another piece is the formalization of \emph{specification notations} for models and algorithms. For example, temporal logics are natural for specifying properties of stateful systems; Hoare logic is a standard mechanism for specifying procedural code; separation logic simplifies the specification of low-level imperative computations; linear logic is suitable for reasoning about resource usage and concurrent behaviors. At this point, there is no unified codebase consisting of such specification notations, and \cslib will address this problem.

\begin{figure}[t]
\begin{mdframed}[roundcorner=10pt]
\begin{minipage}{0.5\textwidth}
(a)
\begin{lstlisting}[basicstyle=\scriptsize\ttfamily]
structure LTS 
    (State : Type u) 
    (Label : Type v) where
  /-- The transition relation. -/
  Tr : State → Label → State → Prop
\end{lstlisting}
(b) 
\begin{lstlisting}[basicstyle=\scriptsize\ttfamily]
def LTS.IsBisimulation
    (lts : LTS State Label)
    (r : State → State → Prop) : Prop :=
  ∀ ⦃s1 s2⦄, r s1 s2 → ∀ μ, (
    (∀ s1', lts.Tr s1 μ s1' → ∃ s2', lts.Tr s2 μ s2' ∧ r s1' s2')
    ∧
    (∀ s2', lts.Tr s2 μ s2' → ∃ s1', lts.Tr s1 μ s1' ∧ r s1' s2')
  )
\end{lstlisting}
\end{minipage}
\begin{minipage}{0.5\textwidth}
\vspace{0.05in}
(c)
\begin{lstlisting}[basicstyle=\scriptsize\ttfamily]
theorem LTS.IsBisimulation.inv
    (h : lts.IsBisimulation r) :
    lts.IsBisimulation (flip r) := by
  simp only [IsBisimulation] at h
  simp only [IsBisimulation]
  intro s1 s2 hrinv μ
  constructor
  case left =>
    intro s1' htr
    specialize h hrinv μ
    have h' := h.2 s1' htr
    obtain ⟨s2', h'⟩ := h'
    exists s2'
  case right =>
    intro s2' htr
    specialize h hrinv μ
    have h' := h.1 s2' htr
    obtain ⟨s1', h'⟩ := h'
    exists s1'
\end{lstlisting}
\end{minipage}
\end{mdframed}
\caption{(a) Definition of a Labeled Transition System (``LTS''), which models the observable behavior of the possible states of a discrete computational system. (b) Definition of a bisimulation between two states. (c) Theorem establishing that the inverse of a bisimulation is a bisimulation.}\label{fig:example}
\end{figure}

\begin{figure}[htbp]
\begin{mdframed}[roundcorner=10pt]
\begin{lstlisting}[basicstyle=\scriptsize\ttfamily]
structure TimeM (α : Type*) where
  ret : α
  time : ℕ

namespace TimeM

protected def pure {α} (a : α) : TimeM α :=
  ⟨a, 0⟩

def bind {α β} (m : TimeM α) (f : α → TimeM β) : TimeM β :=
  let r := f m.ret
  ⟨r.ret, m.time + r.time⟩

instance : Monad TimeM where
  pure := pure
  bind := bind

/-- Advances the time cost by `c` (default 1) without changing the value. -/
def tick (c : ℕ := 1) : TimeM PUnit := ⟨.unit, c⟩

macro "✓[" c:term "]" body:doElem : doElem => `(doElem| do TimeM.tick c; body:doElem)
/-- `✓ x` is `pure x`, adding one tick. -/
macro "✓" body:doElem : doElem => `(doElem| ✓[1] body)

-- Simp lemmas
@[simp] theorem ret_pure {α} (a : α) : (pure a : TimeM α).ret = a := rfl
@[simp] theorem ret_bind {α β} (m : TimeM α) (f : α → TimeM β) :
  (m >>= f).ret = (f m.ret).ret := rfl
@[simp] theorem ret_tick (c : ℕ) : (tick c).ret = () := rfl
@[simp] theorem time_bind {α β} (m : TimeM α) (f : α → TimeM β) :
  (m >>= f).time = m.time + (f m.ret).time := rfl
@[simp] theorem time_pure {α} (a : α) : (pure a : TimeM α).time = 0 := rfl
@[simp] theorem time_tick (c : ℕ) : (tick c).time = c := rfl
-- additional simp lemmas for <* and *> and <*>
(...) 

end TimeM
\end{lstlisting}
\end{mdframed}
\caption{A monadic API for complexity analysis of elementary algorithms. An object of type $\mathtt{TimeM}~\alpha$ comprises $\mathtt{ret}$, which is a value of type $\alpha$, and the $\mathtt{time}$ cost incurred by the computation of this value. The $\mathtt{bind}$ operation chains together two computations and defines the time cost of the composition as the sum of the time costs of the individual operations. The helper function $\mathtt{tick}$ assigns the time cost. The notation $\checkmark$ is syntactic sugar for the invocation of $\mathtt{tick}$. The final section asserts some basic lemmas that help simplify expressions involving $\mathtt{TimeM}$.}\label{fig:complexity-api}
\end{figure}

\begin{figure}[htbp]
\begin{mdframed}[roundcorner=10pt]

\begin{lstlisting}[basicstyle=\scriptsize\ttfamily]

import ComplexityAPI
variable {α : Type} [LinearOrder α]

def merge :  List α → List α  → TimeM (List α)
  | [], ys => return ys
  | xs, [] => return xs
  | x::xs', y::ys' => do
    ✓ let c := x ≤ y
    if c then
      let rest ← merge xs' (y::ys')
      return (x :: rest)
    else
      let rest ← merge (x::xs') ys'
      return (y :: rest)
      
/-- Merge sort algorithm that returns a `TimeM (List α)` where the time represents the total number of comparisons. -/
def mergeSort (xs : List α) : TimeM (List α) :=  do
  if xs.length < 2 then return xs
  else
    let half := xs.length / 2
    let L := xs.take half
    let R := xs.drop half
    let L' ← mergeSort L
    let R' ← mergeSort R
    let result ←  merge L' R'
    return result

/-- MergeSort is functionally correct. -/
theorem mergeSort_correct (xs : List α) :
  IsSorted (mergeSort xs).ret ∧ List.Perm (mergeSort xs).ret xs  := by
  -- Proof omitted
  
/-- Time complexity of mergeSort counted as the number of comparisons. -/
theorem mergeSort_time (xs : List α) :
  let n := xs.length
  (mergeSort xs).time ≤ n * clog2 n := by
  -- Proof omitted
  
\end{lstlisting}
\end{mdframed}
\caption{An implementation of Mergesort using the API in \Cref{fig:complexity-api}. Note how each invocation of the sorting routine returns a sorted list along with a computational cost. With these definitions in place, one can formally state and prove theorems establishing the correctness of Mergesort and its worst-case number of comparisons.}\label{fig:complexity-mergesort}
\end{figure}

While the specific set of formalizations that we will build in this pillar will necessarily evolve over time, 
\Cref{fig:pillar1}-(a) spells out some of the models and logics we currently plan to cover. \Cref{fig:example}-(a) and   \Cref{fig:example}-(b) give concrete examples of formalizations. Specifically, \Cref{fig:example}-(a) shows (part of) a Lean definition of labeled transition systems \cite{winskel1993models}, a classic model for stateful systems, and \Cref{fig:example}-(b) defines bisimulations, a key notion of equivalence between system states. If two states $s_1$ and $s_2$ are related by a bisimulation, then $s_1$ can mimic all transitions of $s_2$ by a corresponding transition and vice versa, and the states reached through these original and mimicking transitions remain related by the bisimulation. 

Transition systems and bisimulations have numerous applications in computer science, in areas from hardware design \cite{chehaibar1996specification} to control \cite{girard2011approximate} to machine learning \cite{zhang2021learning}. Formal definitions of these concepts and associated theorems (such as the theorem in \Cref{fig:example}-(c)) are therefore valuable both from a theoretical and a practical perspective. However, such formalizations are not naturally in scope for Mathlib and are thus an example of the distinct capabilities that \cslib will provide.

\paragraph{Algorithms \& Data Structures.} 
Pillar 1 will also use Lean to build a comprehensive repository of formally verified algorithms and data structures. Our long-term ambition is for this library to cover all of computer science. In the medium term, we aim for this library to cover all algorithms and data structures that a typical computer science graduate is likely to encounter. \Cref{fig:pillar1}-(b) names some of the algorithmic categories that we plan to cover. 

A key objective in this thread is to formalize analysis of algorithms, including, in particular, the \emph{complexity} of algorithms. 
We note that there is much prior work on automatic, lightweight complexity analysis of programs \cite{gulwani2009speed,rosendahl1989automatic}, and some work on proof-assistant-based complexity verification of specific algorithms \cite{chargueraud2019verifying,ChanN21} and formalization in Isabelle~\cite{Nipkow2025FunctionalDataStructures}. However, we are only aware of one effort \cite{irvingdebateframework} on the systematic formalization of algorithmic complexity in Lean. That formalization \cite{brown2023scalable} was focused on stochastic oracle computations; by contrast, we aim for a universal framework for complexity analysis. 

The precise definition of the framework that we propose will require more work and evolve over time. For now, we show in 
\Cref{fig:complexity-api} a simple monadic API, akin to Mathlib's Writer monad and inspired by prior work on analysis of functional algorithms \cite{Danielsson08,GibbonsH11}, that will be part of \cslib's complexity analysis framework. The API packages return values of procedures with the cost of computing these values; 
see \Cref{fig:complexity-mergesort} for an example of its use. 
Treating time complexity as a computational effect makes correctness and complexity analysis orthogonal. 
Also, by giving users control over tick placement, the framework enables experimentation with different cost models, enabling a diverse set of complexity analyses. 

At the same time, the API has multiple limitations. For example, it relies on manual tick annotations that a user could accidentally get wrong rather than automated verification of execution costs. Also, it cannot directly prove statements of the form ``no algorithm can solve problem X faster than $f(n)$." However, because it is lightweight and allows proof of many interesting theorems about algorithmic complexity, it is a good starting point for our efforts. In the longer run, we plan to complement this approach with heavier-weight methods that formalize complexity via explicit RAM and query models.

\subsection{Pillar 2: Infrastructure for Reasoning about Everyday Code}\label{sec:Pillar2}

\cslib's second pillar will develop a powerful new infrastructure for reasoning about everyday imperative code. Lean has recently added substantial support for the definition \cite{ullrich2022unchained} and verification\footnote{\url{https://lean-lang.org/doc/reference/latest/The--mvcgen--tactic/}} of imperative programs, and we intend to leverage these advances in our Pillar-1 efforts. The methods we pursue in the present pillar are complementary in that they connect Lean-based formal reasoning to the verification of code in mainstream languages like Rust, Python, and C++. 

Our approach builds on the rich tradition of deductive verification techniques and tools~\cite{Floyd67,Hoare69}, in particular systems built on top of intermediate verification languages (IVLs). Perhaps best-known is the Boogie IVL~\cite{thisisboogie2,boogieIVL}, which is used by many systems (e.g., Dafny~\cite{leino2010dafny}, Viper~\cite{viper}, CIVL~\cite{civl}, and the Move Prover~\cite{move}).
\cslib's Pillar 2 will be based on a new IVL called \emph{Boole}, partly inspired by Boogie, but also leveraging the capabilities and advantages of Lean.

\begin{figure}[t]
\begin{mdframed}[roundcorner=10pt]
\includegraphics[width=\textwidth]{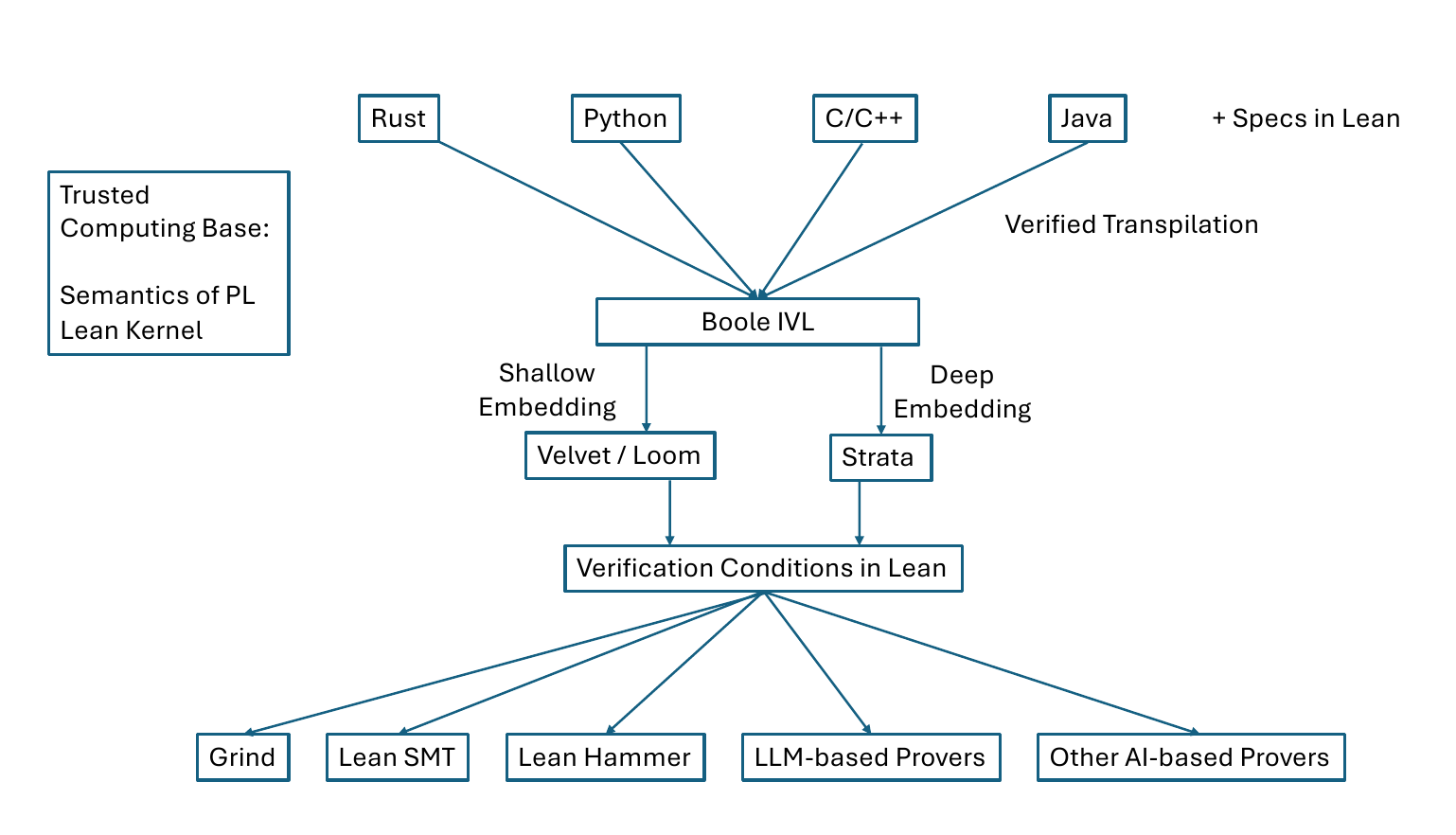}
\end{mdframed}
\caption{CSLib Code Reasoning Vision.}\label{fig:cslib-code}\end{figure}

Figure~\ref{fig:cslib-code} illustrates the long-term vision for \cslib-based code reasoning.  We envision a framework with support for multiple front-end languages.  A user can write their code accompanied by specifications in Lean.  These are then transpiled (translated via compilation) to Boole.  We plan to build a deductive verification platform built around Boole, which can use various algorithms and back-ends to generate verification conditions in Lean.  These can then be discharged using various Lean automation techniques.

\begin{figure}[t]
\begin{mdframed}[roundcorner=10pt]
\begin{minipage}{0.8\textwidth}
\begin{lstlisting}[basicstyle=\scriptsize\ttfamily, keywords={procedure, returns, spec, while, requires, invariant, assert, def, program, var, open, import}]
import Strata.MetaVerifier
import Smt
import Smt.Rat

open Strata

def loopSimple : Strata.Program :=
#strata
program Boole;

procedure loopSimple (n: int) returns (r: int)
spec {
  requires (n >= 0);
}
{
  var sum : int;
  var i : int;

  sum := 0;
  i := 0;
  while(i < n)
    invariant (i <= n && ((i * (i-1)) div 2 == sum));
  {
    sum := sum + i;
    i := i + 1;
  }
  assert [sum_assert]: ((n * (n-1)) div 2 == sum);
  assert [neg_cond]: (i == n);
  r := sum;
};
#end

#prove_vcs loopSimple by
  all_goals (try smt)
\end{lstlisting}
\end{minipage}
\end{mdframed}
\caption{A simple program, written in the Boole dialect of Strata, for computing the sum of the first $n$ positive integers.}\label{fig:sum}
\end{figure}

This vision is guided by several forward-looking design principles, which aim to avoid shortcomings and pitfalls revealed by previous work: ($i$) Boole should look like pseudocode and be easy for humans to understand; ($ii$)
Boole and its ecosystem will be embedded in Lean, and both the transpilation step and the verification condition generation step should be implemented with the goal of minimizing the trusted computing base (TCB); ($iii$) verification conditions should be generated as Lean goals and should be intuitive for humans to read and easily connectable to the code they came from; and ($iv$) code specifications and verification conditions should be able to leverage the rich formal universe defined in \cslib's Pillar 1.

Verification condition generation from Boole will build on infrastructure and ideas from other open source projects, such as Loom~\cite{gladshtein2026foundational} and Strata.  Loom\footnote{\url{https://github.com/verse-lab/loom}} provides a mechanism for the definition of shallowly-embedded languages in Lean (including a Dafny-like language called Velvet), with strong guarantees about correct verification condition generation.  Strata%
\footnote{\url{https://github.com/strata-org/Strata}} is an open-source project that aims at facilitating the creation of domain-specific languages called \emph{dialects}, which are \emph{deeply} embedded in Lean.



 
  

\begin{figure}[t]
\begin{mdframed}[roundcorner=10pt]
\begin{minipage}{0.9\textwidth}
\begin{lstlisting}[basicstyle=\scriptsize\ttfamily]
case entry_invariant_0
⊢ ∀ (n : Int), 0 < n → n ≥ 0 → 0 ≤ n ∧ True

case arbitrary_iter_maintain_invariant_0
⊢ ∀ (n i sum : Int),
  0 < n →
    i < n →
      i ≤ n ∧ i * (i - 1) / 2 = sum →
        n ≥ 0 → i + 1 ≤ n ∧ (i + 1) * (i + 1 - 1) / 2 = sum + i

case sum_assert
⊢ ∀ (n i sum i' sum' : Int),
  n ≥ 0 →
    (if 0 < n then 0 < n else True) →
      (if 0 < n then i < n else True) →
        (if 0 < n then i ≤ n ∧ i * (i - 1) / 2 = sum else True) →
          (if 0 < n then ¬i' < n else True) →
            (if 0 < n then i' ≤ n ∧ i' * (i' - 1) / 2 = sum' else True) →
              (if if 0 < n then False else True then if 0 < n then False else True else True) →
                n * (n - 1) / 2 = if 0 < n then sum' else 0

case neg_cond
⊢ ∀ (n i sum i' sum' : Int),
  n ≥ 0 →
    (if 0 < n then 0 < n else True) →
      (if 0 < n then i < n else True) →
        (if 0 < n then i ≤ n ∧ i * (i - 1) / 2 = sum else True) →
          (if 0 < n then ¬i' < n else True) →
            (if 0 < n then i' ≤ n ∧ i' * (i' - 1) / 2 = sum' else True) →
              (if if 0 < n then False else True then if 0 < n then False else True else True) →
                (if 0 < n then i' else 0) = n∀ (n : Int), 0 < n → n ≥ 0 → 0 ≤ n ∧ True
\end{lstlisting}
\end{minipage}
\end{mdframed}
\caption{Lean verification conditions generated from the program in \Cref{fig:sum}.}\label{fig:goals}
\end{figure}

Figure~\ref{fig:sum} shows a simple snippet of Lean code containing a program written in Boole. Boole is implemented as an extension of the Core dialect of Strata, which in turn is largely modeled after the classic Boogie IVL.  After a prelude that imports and opens relevant packages, the program itself can be written in an easily readable form. Notice the \lstinline{spec} keyword, which signals a specification. In this case, there is a requirement (precondition) that $n$ should be non-negative indicated by the \lstinline{requires} keyword.  We can also specify a loop invariant using the \lstinline{invariant} keyword as well as more general program invariants that should hold at specific points in the code using the \lstinline{assert} keyword.

We can \emph{verify} the assertions by translating the specifications and code into verification conditions.  This can be done by assigning Lean semantics to each construct in the program and then following the standard deductive verification approach, which involves proving, for certain loop-free paths in the program, that for all possible executions along that path, the precondition for the path ensures the postcondition of the path.
The \lstinline{#prove_vcs} command in our example invokes a prototype verification condition generator that generates goals in Lean.  A set of generated goals is shown in Figure~\ref{fig:goals}.  The first goal, for example, represents a path from the entry to the procedure to the beginning of the loop.  The precondition of the program requires $n \ge 0$, and the loop condition must be true, implying  $i < n$.  These should imply the loop invariant.  Our verification condition generator performs some basic simplification---in this case, it knows that $i=0$ when entering the loop, so it replaces $i$ with $0$ and simplifies.  The result is the goal shown.  Similarly, the second goal represents the verification condition stating that the invariant is preserved around one iteration of the loop, and the other two goals correspond to the two assertions.

Goals in Lean can be proved by hand or by invoking \emph{hammers}, which are general-purpose tactics for automating proofs of Lean goals.  A number of potential hammers are shown at the bottom of Figure~\ref{fig:cslib-code}.  For our example, a tactic called \lstinline{smt} can solve all of them.  This tactic uses the Lean-SMT tool \cite{leansmt} to translate a goal into an SMT formula, call a solver, get a proof certificate, and then replay the proof step-by-step in Lean. The number and capabilities of such hammers have been growing rapidly, and the future promises even stronger hammers, including AI-based ones.

The long-term roadmap for \cslib will involve extending Boole over time to add more functionality, including cost semantics (to support reasoning about and proving theorems about computational complexity), support for specifications using definitions from Mathlib and Pillar 1 of \cslib, support for reasoning about concurrency, and support for reasoning about low-level constructs such as pointers.
We will consider modern programming abstractions and verification techniques, for example, as found in research on choreographic programming \cite{Montesi23,RubbensBH24} and separation logic \cite{JungKJBBD18,OHearn2019-cc}.

More significantly, we also intend for Boole to be used as a true \emph{intermediate} language for the verification of real systems in standard programming languages.  The plan is to formalize the semantics of real programming languages in Lean, use those semantics to translate code and specifications in those languages into Boole code and specifications, and then use the verification tools we are building for Boole to verify the translated code.
The Aeneas project \cite{ho2022aeneas}, which translates a sizeable subset of Rust into several proof assistants (including Lean), offers a model of such a pipeline.

A final concrete goal of this pillar is to assemble a large library of code that has been certified using the CSLib verification infrastructure.  We plan to make this library freely available and to integrate it with synergistic efforts such as VeriLib.\footnote{\url{https://verilib.org/}}

\section{The Role of AI}\label{sec:AI}

The \cslib community effort is motivated by the ongoing revolution in AI-based code generation in two ways. On the one hand, we believe \cslib can be a part of our collective response to the \emph{risks} of AI: the ability of AI-powered malicious actors to break the world's software and the possibility that engineers using AI-based coding tools would unwittingly write vulnerable code. These risks make rigorous software verification and formally verified code generation more important than ever, and \cslib will enable this. On the other hand, AI projects such as AlphaProof \cite{alphaproof2025}, Deepseek-Prover \cite{ren2025deepseek}, and Aristotle \cite{achim2025aristotle} showcase the ability of AI to solve hard formal proof tasks. However, AI-based provers are bottlenecked by the abstractions available in the underlying proof assistant. By expanding the universe of CS-relevant abstractions available in Lean, \cslib will also vastly enhance the capabilities of such provers in formal verification settings.

The formalizations produced in \cslib will serve as high-quality training data for AI-based formalization and theorem-proving systems \cite{alphaproof2025,ren2025deepseek,lin2025goedel}. At the same time, we expect that AI-based tooling for Lean proofs will vastly accelerate the development of \cslib. The ideal outcome would be a ``flywheel'' in which both AI-based proof tools and human experts become progressively more efficient at producing new knowledge. 

One risk with the use of AI in a sophisticated development effort like \cslib is that AI-generated formalizations might be buggy. Also, while AI-generated Lean proofs of formal statements may be sound, they may not be human-comprehensible. We will mitigate these risks by ensuring that AI tools play only an advisory role in the development of \cslib, and that all formalizations and proofs that are committed to \cslib are subject to a manual, Github-based code review process.

\section{Building a \cslib Community}

A library is only as impactful as its user base. While formal verification is a vibrant subarea of computer science, we aim for \cslib's impact to transcend that research area. We would like computer science researchers of all sorts, from theorists to system-builders, to use \cslib in their everyday work. We want \cslib to drastically reduce the cost of formal verification to the extent that engineers in a broad range of industries choose to embrace formal techniques. We would like \cslib to be used in a broad swathe of universities to teach the mathematical underpinnings of computer science. 

Fortunately, we can look to Mathlib \cite{baanen2025growingmathlibmaintenancelarge} as a blueprint for achieving our goals. The Mathlib community has demonstrated that open discussion and community activities are essential to making a large-scale formalization project such as ours successful. Following their lead, we have set up a \cslib channel \footnote{\url{https://leanprover.zulipchat.com/\#narrow/channel/513188-CSLib}} on the Lean Zulip chat. 
This channel is already quite active, and we encourage everyone interested in the project to join it. We also urge you to keep an eye out for the calls for contributions to specific components of \cslib that we plan to issue in the coming months. We also plan to hold many workshops and tutorials on \cslib in the coming months and years. If you are interested in joining or leading such an event, please reach out to us! 

Another dimension is the use of rigorous code review practices and the creation of tooling that lowers the cost of \cslib development and maintenance.
In particular, AI tools for automatic formalization and formal theorem-proving have progressed at a breakneck pace over the last few years. As mentioned in \Cref{sec:AI}, a goal of the \cslib effort is to both leverage and improve such tools. We expect these tools to substantially lower the barrier to developing and using \cslib. 

Dependently typed languages have historically existed in a separate universe from mainstream imperative languages, and this is a key reason why their adoption has been limited. \cslib's Pillar 2 aims to change this status quo. The engineering of translations from mainstream languages like Rust and C++ to Boole will be especially important for the success of this vision. We encourage community members to take on the challenge of building these translations. 

Finally, we will prioritize building high-quality, searchable documentation for \cslib. Such documentation will include tutorials targeted specifically at researchers working in different areas of computer science, as well as entry-level textbooks. Please write to us if you would like to lead or contribute to such a textbook.


\section{Governance Model}

\cslib is governed by a dual-body structure designed to balance strategic direction with technical execution. A steering committee comprising leaders from academia and industry is responsible for securing financial support and guiding the project's overall vision. The founding members of this committee are Clark Barrett (Stanford University \& Amazon Web Services), Swarat Chaudhuri (Google DeepMind \& UT Austin), Jim Grundy (Amazon Web Services), Pushmeet Kohli (Google DeepMind), Fabrizio Montesi (University of Southern Denmark), and Leonardo de Moura (Lean FRO \& Amazon Web Services). Working alongside this committee, a maintainer team manages the codebase's technical direction, quality standards, and day-to-day development.
As of the date of this paper, the team is headed by a lead maintainer (Fabrizio Montesi), who coordinates overall efforts, and includes technical leads (Alexandre Rademaker and Sorrachai Yingchareonthawornchai) for long-term cross-cutting developments, and area maintainers (Chris Henson and Kim Morrison) taking ownership of specific domains such as $\lambda$-calculi, metaprogramming, and CI/CD infrastructure.

Naturally, we expect the maintainer team and the steering committee to change as the community grows. Specifically, we plan to periodically invite new maintainers based on merit---particularly, contributions and review activity---and project needs. The goal is to ensure strategic coherence through the steering committee's guidance and technical excellence through the maintainer team's specialized expertise, while fostering a welcoming environment for a broad set of contributors.


\section{Roadmap}

Now we present a brief overview of how we see the \cslib project evolving over 2026 and 2027. This roadmap is tentative given that the project is currently only a few months old and that we plan to refine its scope with community input. A more detailed (and periodically revised) roadmap can be found on the \cslib website: \url{https://cslib.io}.

\paragraph{2026.}  

The \cslib repository already includes initial Lean formalizations of operational semantics, program equivalences, several automata models, a linear logic, and a few elementary sorting and searching algorithms. We expect that by the end of 2026, \cslib will include formalizations of many algorithms covered in a typical undergraduate algorithms and data structures course, and most of the models and logics covered in a typical undergraduate theory of computation course. In addition, community members will be welcome to contribute formalizations from specialized research areas that they work on. Note that progress on this Pillar-1 task does not depend on Boole to be ready.

As for Pillar 2, we have already developed an initial version of Boole on top of the Strata framework. Over the next year, we expect the framework to mature substantially. By the end of 2026, the Boole framework will have robust capabilities for generating Lean-language verification conditions and interacting with SMT-based hammers. By that point, the \cslib repository will include Boole-language representations of a sizeable number of elementary algorithms and data structures, along with Lean specifications and proofs of all relevant verification conditions. 

\paragraph{2027.}

The project's second year will focus on scale and unification. In particular, our Pillar-1 efforts will move on to technically challenging topics such as complexity theory, concurrency, secure compilation, and randomized and quantum algorithms. We expect that as a result, \cslib will have substantially aided at least one important foundational discovery in computer science by the end of 2027. In our Pillar-2 efforts, we will extend Boole to support machinery, such as separation logic, that is needed for the verification of low-level systems code, and extensively use the Pillar-1 machinery to write more advanced system specifications. We expect that by the end of 2027, \cslib will have enabled the end-to-end verification of at least one substantially sized real-world system.

\subsection*{Acknowledgements}

We thank all current and future contributors to \cslib \footnote{A list of current \cslib contributors is available at \url{https://github.com/leanprover/cslib/graphs/contributors}.} for their work on the project, and Eric Wieser and Tom Kalil for their thoughtful feedback on this paper. 

\bibliographystyle{abbrv}
\bibliography{refs}  

\end{document}